\documentclass[article]{elsarticle}

\usepackage{lineno}
\modulolinenumbers[10]

\journal{Annals of Physics}

\usepackage{graphicx}
\usepackage{ulem}
\usepackage{soul}
\usepackage{amsmath}
\usepackage[colorlinks, linkcolor=blue]{hyperref}
\def \k {{\bf k}}
\def \a {\alpha}
\def \p {{\bf p}}

\def \ve {\varepsilon}
\def \r {{\bf r}}

\def \el {\ell}
\def \s {\sigma}
\def \q {{\bf q}}

\def \Q{{\bf Q}}
\def \l {\ell}
\def \ve {\varepsilon}

\def \S {{\cal{S}}}

\def \x {{\bf x}}

\def \beq {\begin{eqnarray}}
\def \eeq {\end{eqnarray}}
\def \tn {\textnormal}

\def \ua {\uparrow}
\def \da {\downarrow}
\def \ie {{\it i.e.}}
\def \eg {{\it e.g.}}
\begin{document}

\begin{frontmatter}
\title{The unreasonable effectiveness of Eliashberg theory for pairing of non-Fermi liquids\tnoteref{rev}}
\tnotetext[rev]{Invited review for special issue dedicated to Gerasim Eliashberg. \\Title adapted from ``{\it The unreasonable effectiveness of mathematics in the natural sciences}", E.P. Wigner, Math. $\&$ Sci., 291-306 (1990).}

\author{Debanjan Chowdhury}
\address{Department of Physics, Cornell University, Ithaca, New York 14853, USA.}
\ead{debanjanchowdhury@cornell.edu}

\author{Erez Berg}
\address{Department of Condensed Matter Physics, Weizmann Institute of Science, Rehovot, 7610001, Israel.}
\ead{erez.berg@weizmann.ac.il}

\begin{abstract}
The paradigmatic Migdal-Eliashberg theory of the electron-phonon problem is central to the understanding of superconductivity in conventional metals. 
This powerful framework is justified by the smallness of the Debye frequency relative to the Fermi energy, and allows an enormous simplification of the full many-body problem. However, superconductivity is found also in many families of strongly-correlated materials, in which there is no {\it a priori} justification for the applicability of Eliashberg theory. In these systems, superconductivity emerges out of an anomalous metallic state, calling for a new theoretical framework to describe pairing out of a non-Fermi liquid. 
In this article, we review two model systems in which such behavior is found: a Fermi sea coupled to gapless bosonic fluctuations, and a system of fermions with local, strongly frustrated interactions. In both models, there is a well-defined limit in which the Eliashberg equations are asymptotically exact even in the strongly coupled regime. These models thus provide tractable examples of how superconductivity can emerge in the absence of coherent electronic quasiparticles; they also demonstrate the surprisingly wide applicability of the Eliashberg formalism, well beyond the conventional regime for which it was originally designed.

\end{abstract}


\end{frontmatter}

\tableofcontents

\section{Introduction}
\label{intro}

Superconductivity (SC) is often regarded as one of the most striking examples of a macroscopic collective quantum phenomena in a many-electron system. Since the original discovery of SC in mercury in 1911, it took nearly fifty years and the trio of  Bardeen-Cooper-Schrieffer (BCS) to arrive at a microscopic theory of phonon mediated superconductivity in conventional metals~\cite{BCS}. 
The key conceptual framework that led to the foundation of a strong-coupling theory for electron-phonon superconductivity is due to Migdal~\cite{migdal} and Eliashberg~\cite{eliashberg}. This theory has enjoyed tremendous successes, explaining countless experiments on conventional superconductors, and even predicting new ones. A remarkable recent example is the prediction and subsequent discovery of near-room temperature superconductivity under extremely high pressure in metallic hydrides \cite{PThyd}.

Remarkably, Migdal-Eliashberg (ME) theory is controlled in a well-defined limit, which does not necessarily rely on the smallness of the dimensionless electron-phonon coupling $\lambda$, but rather on a small $(\lambda \theta_D/\ve_F)$, where $\theta_D$ and $\ve_F$ are the Debye temperature and Fermi energy, respectively. In this limit, Migdal and Eliashberg showed that the full many-body problem can be reduced to solving a set of non-linear self-consistent equations. These equations capture a host of physical effects, including the retardation of the phonon-mediated electron-phonon interaction, its interplay with the bare Coulomb repulsion, and the renormalization of both the electronic and phononic quasi-particles; see Ref.~\cite{DJSeliash} for a detailed exposition. In particular, ME theory explains how, thanks to the effects of retardation, $T_c$ can be non-zero even if the bare Coulomb repulsion is stronger than the bare phonon-mediated attraction.{\footnote{A natural follow-up question is if there is a bound on how large $T_c$ can be, which has recently been addressed \cite{Esterlis18a,Esterlis18b}. It is clear that for a sufficiently large $\lambda$, polaron formation, not captured in ME theory, will ultimately suppress superconductivity.}} 

For electron-phonon mediated spin-singlet superconductivity in `good' metals ($\ie$ those which host a sharp Fermi surface in the limit of $T\rightarrow0$, and have well-defined low-energy quasiparticles), $T_c$ is ultimately determined by the tendency of the quasiparticle states near a filled Fermi sea at $|\k\uparrow\rangle$ and $|-\k\downarrow\rangle$ to form a bound state---this is the celebrated `Cooper-problem' \cite{cooper}. In contrast, one of the recurring themes in the study of `strongly-correlated materials' in the last few decades has been the emergence of superconductivity in compounds where the parent metallic state has highly anomalous properties that are at odds with the expectations in a Landau Fermi-liquid (FL). In these systems there is ample empirical evidence that the pairing is not due to a purely phonon-based mechanism. The most well studied compounds include the copper-oxide based (``cuprate") \cite{Keimer15}, iron-pnictide (chalcogenide) based \cite{scalapino} and certain rare-earth element based \cite{Stewart} compounds, where some of the peculiarities include, $\eg$ uncharacteristically short single-particle lifetimes \cite{zxs,johnson} and a broad regime of anomalous non-Fermi liquid (NFL) power-law transport \cite{Takagi,Marel,Taillefer1,hussey} amongst others. One of the striking features is the presence of an underlying sharp Fermi surface, where the low-energy quasiparticles are not long-lived even arbitrarily close to the Fermi surface.

One of the holy grails in the field is the nature of pairing instabilities out of such NFL states that host a {\it critical} Fermi surface (CFS)---a sharp electronic Fermi surface without any low-energy Landau quasiparticles. There are a number of pertinent questions, which include: (i) Is there an underlying Cooper instability for a CFS and is it an essential ingredient for superconductivity? (ii) Can a CFS be unstable to superconductivity as a result of the same electronic interactions that lead to the destruction of the quasiparticles in the first place? (iii) Are there generic `intertwined' ordering tendencies that compete with superconductivity in the regime of strong interactions? (iv) What is the role played by the Fermi surface geometry (`fermiology') in determining the nature of these instabilities?

The answers to many of these questions in models with purely electronic interactions remain poorly understood for the most part. Finding concrete examples of electronic models, either defined on the lattice or in the continuum (as effective field theories), where the emergence of NFL behavior and superconductivity can be analyzed through reliable theoretical means is challenging and remains of paramount importance. In this short article, we review some of the recent understanding of superconductivity in two classes of electronic models. We begin by discussing the low-energy effective field theory for a FL coupled to the fluctuations of a gapless bosonic field, that can arise, e.g., at a quantum critical point (QCP) to some form of broken symmetry, or for (non-local) fermions that are charged under an emergent gauge field. Next we discuss a solvable example of a lattice electronic model with frustrated interactions. In both of these two cases, there is a well defined limit where a NFL regime emerges. 
Since there is no separation of scales between the Fermi energy and the characteristic energy scale of the bosonic fluctuations in the problem, a Migdal type argument does not apply. 
Nevertheless, in both problems there is a well-defined theoretical limit where  Eliashberg theory becomes exact, and captures the emergence of a non-Fermi liquid state and its interplay with superconducting instabilities, or lack thereof. We focus on some of the recent conceptual advances, addressing specifically the questions raised above.

The rest of this article is organized as follows. In Sec.~\ref{qcmetal}, we review the low-energy field theoretic formulation for a Fermi surface coupled to the fluctuations of a gapless bosonic field. We focus specifically on a controlled setup to treat the problem at strong-coupling in Sec.~\ref{double} and consider the pairing instabilities of the CFS thus obtained in Sec.~\ref{pairQC}. In Sec.~\ref{syk}, we introduce lattice models with the hope of describing non-Fermi liquids at strong coupling. We review a relatively recent approach that utilizes a solvable limit in Sec.~\ref{syksol} to describe different non-Fermi liquid regimes, including an incoherent metal and a marginal Fermi liquid (Sec.~\ref{mfl}). In Sec.~\ref{SCsyk}, we describe the instabilities of these states to pairing and other forms of order. We summarize these results and discuss them in a broader context in Sec.~\ref{outlook}.

\section{Non-Fermi liquid from bosonic quantum criticality}
\label{qcmetal}

In this section, we review the classic problem of a Landau Fermi liquid coupled to the fluctuations of a collective mode, $\ie$ a bosonic field. The framework often goes under the name of Hertz-Millis-Moriya (HMM) criticality \cite{Hertz,Millis,moriya}. When the bosonic field is gapless, its low-energy critical fluctuations can destroy the Fermi liquid behavior;  the effect is particularly strong in two-dimensional systems, as we discuss below. This problem has appeared in a large number of different settings in the study of strongly correlated systems. Broadly speaking, the bosonic field can be classified on the basis of whether it carries zero or non-zero momentum, $\Q=0$ or $\Q\neq0$. 

The former example ($\ie$ $\Q=0$ boson) includes the problem of a Fermi surface coupled to a gapless transverse gauge boson, which appears as a low-energy description of a U(1) spin-liquid with a spinon Fermi surface \cite{SFS}, the Halperin-Lee-Read (HLR) theory for the half-filled Landau level \cite{HLR}, and other related non-Fermi liquid phases. In all of these above examples, the boson is naturally gapless. A completely different example, that can be described within the same framework, is the case of a Fermi liquid at the brink of a Pomeranchuk instability to broken rotational symmetry \cite{nemrev}. Specifically, for an interacting Fermi liquid at the onset of an electronic nematic order, the critical point is described by the same low-energy action as detailed in the section below, where the only difference is in the precise form of coupling between the fermions near the Fermi surface and the bosonic field \cite{MMSS10a}. 

In all of the above setups, it is important to address the fate of the infra-red fixed point structure of the theory, if there exists one. In particular, does the Fermi surface survive down to zero temperature realizing a $T=0$ NFL with a sharp Fermi surface, or is there a pre-emptive instability to pairing (possibly even other competing orders) due to the interactions mediated by the same gapless boson? As it turns out, the answer to this question depends on the underlying microscopic details, as we shall review below. 

The example with $\Q\neq0$ describes the problem of an onset of a `density-wave' ($\ie$ a spin, charge or pair-density wave) in a Landau Fermi liquid. In the ordered-phase, as long as $|\Q|<2|\k_F|$, 
the Fermi-surface gets reconstructed where the gap opens up near a set of points (in two spatial dimensions) referred to as `hot-spots'. At the critical point, the gapless boson leads to an enhanced scattering near these hot-spots, where NFL behavior sets in; regions far away from these points retain their FL-like behavior. We will not discuss the subtle interplay between the hot and cold regions along the Fermi surface, and how they interact with the gapless boson (as well as various composite operators \cite{MMSS11}), to give rise to pairing in this review; this has been discussed elsewhere \cite{AVCJSrev,AJM92,AMT03,MMSS10b,AVC16,EB17}. Instead, here, we focus on the case where the bosonic fluctuations are gapless at $\Q=0$, and as a result, the entire Fermi surface (with the possible exception of a discrete set of `cold spots') is strongly coupled to the gapless boson. 

\subsection{Controlled expansion at strong-coupling}
\label{double}

We are interested in describing the low-energy physics of the fermionic states near the Fermi surface at $\k=\k_F$ coupled to the long-wavelength/low-frequency modes of the gapless $\Q=0$ boson. The minimal Euclidean action is given by,
\beq
\S &=& \S_\psi + \S_a + \S_{\tn{int}},\label{eq:S}\\
\S_\psi &=& \int_{\k,\omega} \psi^\dagger_{\k\a} (-i\omega + \ve_\k - \mu)\psi_{\k\a},\\
\S_a &=& \frac{1}{e^2}\int_{\k,\omega} \k^2 |a(\k,\omega)|^2 \label{sa},\\
\S_{\tn{int}} &=& \int_{\k,\omega} a(\k,\omega)~O(-\k,-\omega),
\eeq
where $\psi_{\k\a}$ with $\a=1,...,N$ represents an $N$ component fermion ($N=2$ in most cases of physical interest for spinful fermions) and $a$ is the gapless boson. Depending on the details of the specific underlying problem, $a$ is either the transverse component of a gauge-field, or the nematic order parameter. In the former example, $O(\x,\tau)$ represents the transverse current density for the $\psi$ fermions, while in the latter it is the $\psi-$bilinear that transforms with the same symmetry as the order parameter.

\begin{figure}[t]
	\begin{center}
		\includegraphics[width=0.80\columnwidth]{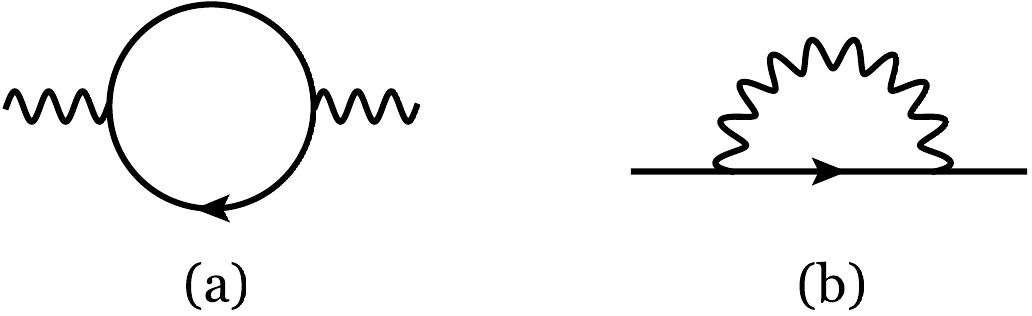}
	\end{center}
	\caption{The (a) bosonic self-energy, $\Pi(\q,\Omega)$, and, (b) fermionic self-energy, $\Sigma(\k,\omega)$. Solid (wiggly) lines denote fully dressed fermionic (bosonic) propagators.}
	\label{rpa}
\end{figure}
	
There is a long history of treating the problem within random-phase approximation (RPA), which can be systematically organized within a $1/N-$expansion (Fig.~\ref{rpa}). The basic idea is to first generate a Landau-damping term for $a(\q,\Omega)$ (in the limit of $\Omega\ll v_F q$) as a result of coupling to the gapless Fermi surface. Upon scattering off these Landau-damped (gapless) bosons, the self-energy for the fermions becomes singular and leads to a short lifetime. Focusing specifically on the case of two-dimensions, such an RPA treatment leads to,
\beq
D(\q,\Omega) &=& \frac{1}{\q^2/e^2 - \Pi(\q,\Omega)},\\
\Pi(\q,\Omega) &=& \nu\bigg(1 - \frac{|\Omega|}{\sqrt{\Omega^2 + (v_Fq)^2}} \bigg),\\
\Sigma(\k_F,\omega) &=& -\frac{e^{4/3}}{N}\frac{i~\tn{sgn}(\omega)}{(\nu v_F^2)^{1/3}} |\omega|^{2/3}, 
\eeq
 where $D$, $\Pi$, and $\Sigma$ are the boson propagator, boson self-energy and fermion self-energy, respectively, $\nu$ is the density of states near the Fermi energy, and $v_F$ is the Fermi velocity (see Fig.~\ref{rpa}). Thus, the fermionic self-energy is more singular than the bare `$i\omega$'-term, resulting in a NFL behavior. In fact, at large$-N$, the fully self-consistent set of equations for the fermion and boson Green's functions in terms of the dressed propagators (and ignoring the vertex renormalizations) leads to the above results.  

Unfortunately, the above large$-N$ expansion suffers from a problem and does not help control the theory in two spatial dimensions \cite{SSL09}. Naively, the advantage of setting up the $1/N$ expansion is that a higher loop diagram appears to be higher order in this expansion. However, the loop integrals for a large subset of these diagrams have divergences which can be cured by including the $1/N$ (one-loop) fermion self-energy. This leads to a trade-off where the singular divergence, upon being cured, leads to an $N-$dependence in the numerator, thereby modifying the entire counting. Formally, it was pointed out \cite{SSL09} that for the a theory with a single Fermi surface ``patch'', there is a set of planar diagrams that can be organized systematically and the self-energy takes the form,
\beq
\Sigma(\k_F,\omega) \propto i~\tn{sgn}(\omega) |\omega|^{2/3}\sum_m\frac{b_m N^{m-1}}{N^m}.
\eeq
In the absence of the subtleties explained above, formally the expansion would only be of order $1/N^m$ at each order, $m$. However, controlling the divergence leads to the said enhancement by the factor of $N^{m-1}$. The resulting low-energy structure of the theory thus becomes far more complicated than can be reliably treated within the simple $1/N$ expansion and the eventual fate of the possibly singular nature of the diagrams remains unclear.

An alternative way in which the theory can be made controlled is to make it non-local by a `small' amount \cite{nayak,AIM94,mross}, $\ie$ to generalize $\S_a$ in Eq.~\ref{sa} as,
\beq
\S_a = \frac{1}{e^2} \int_{\k,\omega} |\k|^{z_b-1} |a(\k,\omega)|^2,
\label{planarse}
\eeq
and study the problem in the limit of a  small $\epsilon~ (= z_b - 2)\ll1$; thus $\epsilon=0$ ($\epsilon=1$) corresponds to the HLR problem with Coulomb interactions (spinon Fermi surface and nematic QCP). How does this modification provide us with a controlled route towards describing a NFL regime, and more importantly, when is it legitimate to introduce a small parameter to vary the dynamical exponent? 

To begin with, for the one-patch theory, the modified expression for the self-energy after including the sum over the same set of planar diagrams is of the form \cite{mross},
\beq
\Sigma(\k_F,\omega) \propto i~\tn{sgn}(\omega) |\omega|^{2/z_b}\sum_m\frac{c_m (\epsilon N)^{m-1}}{N^m}.
\eeq 
The advantage of this setup is that as long as $\epsilon~ (=z_b-2)\sim 1/N$, the above expression becomes controlled in the $N\rightarrow\infty$ limit and is dominated by the first term in the expansion ($\ie~m=1$, which is essentially the one-loop result). As was argued earlier \cite{nayak,mross} and is clear on physical grounds, such a prescription should be effective when the exponent is not expected to renormalize under a Wilsonian RG. Thus, once the dust settles, there is a well-defined way of controlling the expansion in a systematic fashion at the cost of making the bosonic action infinitesimally non-local. The resulting NFL has a sharp critical Fermi surface but the low-energy excitations are not Landau quasiparticles{\footnote{To be clear, for the problem of the spinon Fermi surface, the spinons are non-local fractionalized quasiparticles and are not `Landau' quasiparticles' to begin with; as a result of the coupling to the gauge-field even the spinons are no longer well-defined quasiparticle excitations. }}---they have an anomalously short lifetime controlled by the exponent, $z_b$.

The above expansion scheme has the appealing feature that it can describe different classes of NFL regimes (depending on the physical value of $\epsilon$) that emerge from distinct microscopic settings. Given that the low-energy theories for the Fermi surface coupled to the nematic order-parameter vs. gauge field look so similar{\footnote{As a reminder, the former couples to the electronic Fermi surface while the latter couples only to an electrically neutral spinon Fermi surface.}}, we may ask if there are any differences as far as the various low-energy response functions are concerned? Moreover, from the discussion so far, it appears that the above presciption has allowed one to describe a NFL regime that is stable down to $T\rightarrow0$, $\ie$ it is `infra-red complete'. However, we have not addressed the stability of such NFLs to other ground-states, including superconductivity, thus far. As will become clear shortly, these two questions are intimately tied together and will lead us to address how pairing may or may not emerge in these two cases.

It turns out that there is an important distinction, tied to the ``Amperean'' interaction between two fermions mediated by the boson, in the two cases. For the gauge-field problem, the particle current associated with one patch is parallel to the hole current in the antipodal patch; the gauge-mediated interaction for the ``$2K_f$'' particle-hole pair is therefore attractive. On the other hand, the currents for the fermions, but now with one from either patch are antiparallel and the resulting gauge-mediated interaction is repulsive in the particle-particle Cooper channel. Due to the precise form of the Yukawa coupling in both the cases, the situation is exactly reversed for the nematic problem. Thus, it is not difficult to imagine that the gauge-field fluctuations suppress tendency towards an instability in the Cooper channel but the nematic field enhances the same tendency. We discuss this in a little more detail in the next section.

\subsection{Pairing instabilities}
\label{pairQC}

\begin{figure}[t]
	\begin{center}
		\includegraphics[width=0.95\columnwidth]{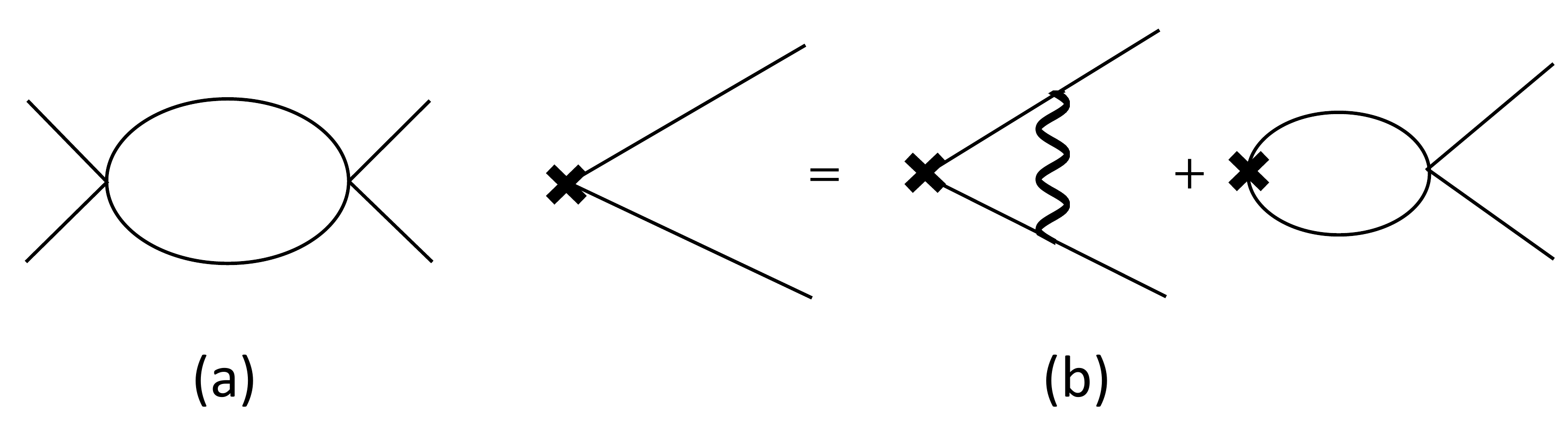}
	\end{center}
	\caption{(a) Renormalization of the BCS interaction in a Fermi liquid, and, (b) Linearized gap equation for the pairing vertex (solid cross) due to exchange of gapless bosonic fluctuation and four-fermion interaction. The solid and wiggly lines represent fully dressed fermion and boson propagators, as in Fig.~\ref{rpa}.}
	\label{eliashnem}
\end{figure}

Before focusing our attention on quantum critical metals, let us first briefly review the problem of pairing in conventional metals. As is well known for ordinary Fermi liquids, forward and BCS scattering are the only kinematic processes that survive after carrying out a Wilsonian RG transformation \cite{shankar,polchinski}. The four-fermion interaction in the BCS channel can be expressed as,
\beq
&&S_{\tn{BCS}} = \nonumber\\
&&-\frac{1}{4}\prod_{i=1}^4\int_{\k_i,\omega_i} \psi^\dagger_{\k_1\alpha}\psi^\dagger_{\k_2\beta}\psi_{\k_3\gamma}\psi_{\k_4\delta}~\delta^2(\k_1+\k_2-\k_3-\k_4)~\delta(\omega_1+\omega_2-\omega_3-\omega_4)~\nonumber\\
&&\bigg[ (\delta_{\alpha\gamma}\delta_{\beta\delta} + \delta_{\alpha\delta}\delta_{\beta\gamma})~V^a(\k_1,\k_2;\k_3,\k_4) + (\delta_{\alpha\gamma}\delta_{\beta\delta} - \delta_{\alpha\delta}\delta_{\beta\gamma})~V^s(\k_1,\k_2;\k_3,\k_4)\bigg],\nonumber\\
\label{eq:BCS}
\eeq
where $V^{s(a)}$ represent functions of momenta that are symmetric (antisymmetric) under exchanging $\k_1\leftrightarrow\k_2$, $\k_3\leftrightarrow\k_4$. Focusing specifically on the BCS case and superconducting solutions at zero center of mass momentum, we choose $\k_2=-\k_1$, $\k_3=-\k_4$. Restricting these momenta to lie on the FS for a rotationally invariant system, these coupling constants can be simplified to $V^{s(a)}(\k_1,-\k_1;\k_2,-\k_2) = V^{s(a)}(\theta_1-\theta_2) = \sum_m V^{s(a)}_m e^{im(\theta_1-\theta_2)}$, where we have expanded the interaction in terms of all the allowed harmonics. Upon carrying out the usual Wilsonian RG for a FL \cite{shankar,polchinski}, the above interaction turns out to be marginal and has the following flow at one-loop (see Fig. \ref{eliashnem}a)
\beq
\frac{dV_m^{s(a)}}{d\el} = - (V_m^{s(a)})^2,
\label{FLbcs}
\eeq
where the coupling constants have been rescaled by a factor of $(k_F/v_F)$ to make them dimensionless. Clearly, if we start with a repulsive $V_m^{s(a)}>0$, it flows logarithmically to zero. On the other hand, an attractive interaction $V_m^{s(a)}<0$ leads to a runaway flow and an instability to pairing at a scale set by $\Lambda~\tn{exp}(-1/|V_m^{s(a)}|)$, which is the usual BCS instability to superconductivity ($\Lambda\equiv$ultraviolet cutoff). How does the coupling to the gapless boson modify the above picture?

In order to address this question within the two-patch formalism, it is useful to first define the dimensionless coupling constant, $\alpha\equiv e^2v_F\Lambda^{-\epsilon}/(2\pi)^2$. Within a one-loop RG, the coupling has the flow \cite{nayak,mross}
\beq
\frac{d\alpha}{d\ell} = \frac{\epsilon}{2}\alpha - \frac{\alpha^2}{N}.
\label{alpharg}
\eeq
Thus, depending on whether $\epsilon=0$ or $\epsilon>0$, $\alpha$ either flows logarithmically to zero or to a fixed point value, $\alpha^*=\epsilon N/2$. We now have an interesting situation where naively one expects the gapless fluctuations of the boson (which mediate long-range interactions between the fermions) to enhance the tendency towards pairing while the same boson also destroys the quasiparticles near the Fermi surface. The key question boils down to which of these two effects dominates.

To examine this question, we add a local interaction of the form of Eq.~\ref{eq:BCS} to the action of the critical metal, Eq.~\ref{eq:S}. 
In the large$-N$ and weak $V_m$ limit, the primary effect of the gapless boson is to renormalize the effective BCS interaction. This is shown as the one-boson exchange process in the pairing vertex in Fig.~\ref{eliashnem}b and upon integrating out over the bosonic modes leads to an inter-patch four-fermion interaction,
\beq
\delta V^{s(a)}(\k_1,-\k_1;\k_2,-\k_2) = -\frac{\zeta}{2} v_F^2 D_{>}(0,\k_1-\k_2),
\eeq
where $\zeta = +1~(-1)$ for the nematic (gauge-field) problem and $D_>(\omega,\k)$ denotes the bosonic propagator which includes contributions from `high-energy' modes, $\Lambda~e^{-\ell/2}<|\k|<\Lambda$. The RG flow for the BCS coupling thus becomes \cite{MM15},
\beq
\frac{dV_m^{s(a)}}{d\el} = -\zeta \frac{\alpha}{N} - (V_m^{s(a)})^2,
\label{nflbcs}
\eeq
where the first term, which is independent of $m$, arises from the gapless boson (and has an interesting sign-structure) while the second term is the same as Eq.~\ref{FLbcs}. The combination of Eqs.~\ref{alpharg} and \ref{nflbcs} determines the fate of the NFL and its potential instability to pairing.

For the problem of a Fermi surface coupled to a nematic order parameter ($\zeta=+1$), where the gapless boson mediated interactions in the pairing channel become attractive, the above analysis leads to an interesting result. In the regime where the calculation is controlled ($\ie$ when $\epsilon\ll 1$), if one compares the scale ($\Lambda_{\tn{NFL}}$) at which the NFL behavior sets in and the quasiparticles get destroyed with the pairing scale ($\Delta_{\tn{pair}}$), one finds $\Lambda_{\tn{NFL}}\ll\Delta_{\tn{pair}}$ \cite{MM15}. Thus, the pairing instability preempts the onset of non-Fermi liquid behavior and any nematic quantum criticality in the limit of $T=0$ necessarily occurs deep inside the superconducting dome. However, as one takes the physical limit of $\epsilon\rightarrow1$, the above two scales approach other and the problem becomes intractable once again. It is worth noting that for most of the material families where superconductivity appears out of a `mother' NFL state, the NFL regime typically extends to scales that are significantly higher than the scale of $T_c$. Finally, we note that the above result is reminiscent of color superconductivity of baryonic matter in three spatial dimensions \cite{DTS} and pairing near a ferromagnetic QCP in a metal in three spatial dimensions \cite{AVCJS}.

The problem of a metal near a nematic quantum critical point is free of the sign problem, thanks to the presence of two spin flavors. This problem has been studied extensively~\cite{Schattner2016,Lederer2017,berg2019monte} using the determinant quantum Monte Carlo technique. In particular, it has been found that the quantum critical point is covered by a broad superconducting `dome'. Moreover, in all the cases that have been studied, superconductivity onsets near (or slightly below) `non-Fermi liquid' temperature scale, defined as the point where the maximal fermion self energy on the Fermi surface exceeds the bare $i\omega$ term in the inverse fermion propagator. Thus, it seems that in the nematic problem with $\epsilon=1$ and $N=O(1)$, superconductivity indeed preempts the development of a full-fledged NFL, in agreement with the arguments above. Despite this, the transport properties display anomalous, non-Fermi liquid behavior{\footnote{These results are based on certain ``resistivity proxies", obtained from an analytic continuation of imaginary time Monte Carlo data.}} in a broad regime of temperatures above $T_c$~\cite{Lederer2017}. 

Let us now return to the problem of Fermi surface coupled to a gauge field ($\zeta=-1$). In contrast to the nematic problem, the gauge field fluctuations now drive the interaction in the pairing channel repulsive. For $\epsilon$ small and positive (corresponding to the problems of spinon Fermi surface and HLR with short-ranged interactions), the fixed-point structure from Eqs.~\ref{alpharg} and \ref{nflbcs} leads to $\alpha^*=N\epsilon/2$ and $V_\pm^*=\pm\sqrt{\epsilon/2}$. The fixed point ($\alpha^*,~V_+^*$) is stable in the infrared. Thus the spinon Fermi surface and HLR phase are stable against pairing as long as the initial value of $V$ is greater than $V^*_-$ \cite{MM15}. Most importantly, a critical initial value of the coupling is needed to drive a pairing transition (to a $Z_2$ spin liquid from the U(1) spin liquid or to an incompressible quantum Hall insulator from the HLR phase) out of the NFL. Thus, unlike the case of the nematic problem, the `mother' NFL regime is not masked by a tendency of fermions to pair;  a critical coupling strength is needed to compensate for the long-range repulsive interaction mediated by the gauge-field. Finally, we note that for $\epsilon=0$, corresponding to HLR with long-ranged Coulomb interactions, the two fixed points $V^*_\pm$ found above merge into a single fixed point and as long as $V>-\sqrt{\alpha}$, the HLR phase remains stable \cite{MM15}. On the other hand, if pairing does occur due to interactions generated from short-distance physics, the resulting transition in the $(p+ip)-$channel into the incompressible Moore-Read phase \cite{MR} can be continuous{\footnote{This is consistent with some of the previous numerical studies \cite{Haldane00,Cooper11,Haldane12}.}}. {Finally, we note in passing that some of the above theoretical considerations might be phenomenologically relevant in the context of frustrated Mott insulators in certain triangular lattice organic compounds; see Ref. \cite{KKrev} for a discussion.}

{It is important to note that we have focused primarily on conventional BCS pairing of spinons with zero center of mass momentum in this review. To the best of our knowledge, within the $\epsilon-$expansion, the spinon fermi surface is stable against pairing towards more exotic states with a finite center of mass momentum, such as the one considered in Ref.~\cite{PAL07}. This does not, however, rule out the possibility of such an exotic paired state to arise out of the spinon FS state at strong-coupling. Similarly, depending on the precise microscopic details of the short-ranged interaction between the spinons, it is possible to have other exotic paired states (e.g., Ref.~\cite{VG07}), which we have not focused on here for reasons of simplicity.} 

The RG analysis of the pairing instabilities, controlled within the double expansion in $\epsilon$ and $1/N$, is identical to studying the pairing vertex within the Eliashberg approximation~\cite{MM}; see Fig. \ref{eliashnem}b. Thus, the problem of a Fermi sea coupled to a gapless boson at ${\bf Q}=0$ within the above controlled framework is a remarkable example of the applicability of Eliashberg theory, well beyond its original intended use\footnote{Eliashberg theory has been applied to the problem of pairing of the HLR state in past~\cite{Nayak1996,Wang_2014,Isobe2017}, although with no formal justification.}.

\section{Non-Fermi liquid from frustrated interactions}
\label{syk}

In the previous section, we discussed a concrete example of a NFL and its propensity towards a superconducting instability within a low-energy field theoretic approach. In this section, we present a departure from this paradigm and instead focus on a complementary framework constructing lattice models for NFLs. We consider multi-orbital electronic models with a single globally conserved density and interacting with strong local SU(2) invariant interactions. Let us start with a generic model defined on the sites, $\r$, of a hypercubic lattice in an arbitrary number of dimensions{\footnote{We will be primarily interested in $(2+1)-$dimensions, but our constructions work in any number of dimensions.}} (lattice spacing $\equiv a$), with $N-$orbitals at every lattice site labelled by $i=1,..,N$ and spin $\s=\ua,\da$,
\beq
H_d &=& H_{\tn{kin,d}} + H_{\tn{int}},\\
H_{\tn{kin,d}} &=& \sum_{\r,\r'}\sum_{\substack{i\\ \s=\ua,\da}}(-t_{\r\r'} - \mu_d\delta_{\r\r'}) d^\dagger_{\r i\s}d_{\r' i\s} \\
H_{\tn{int}} &=& \frac{1}{N^{3/2}}\sum_{\r}\sum_{\substack{i,j,k,\l\\ \s,\s'=\ua,\da}} U_{ijkl}d^\dagger_{\r i\s}d^\dagger_{\r j\s'}d_{\r k\s'}d_{\r \l\s}.
\label{oneband}
\eeq
The hopping parameters are denoted, $t_{\r\r'}$, the chemical potential, $\mu_d$, can be tuned to adjust the total conserved $U(1)$ density for the $d-$electrons and $U_{ijk\l}$ represent the on-site interaction matrix-elements.

It is useful to consider an additional multi-orbital system of $c-$electrons (with an independent globally conserved density) coupled to the $d-$electrons as,
\beq
H_c &=& H_{\tn{kin,c}} + H_{\tn{cd}},\\
H_{\tn{kin,c}} &=& \sum_{\r,\r'}\sum_{\substack{i\\ \s=\ua,\da}}(-\zeta_{\r\r'} - \mu_c\delta_{\r\r'}) c^\dagger_{\r i\s}c_{\r' i\s} \\
H_{\tn{cd}} &=& \frac{1}{N^{3/2}}\sum_{\r}\sum_{\substack{i,j,k,\l\\ \s,\s'=\ua,\da}} V_{ijkl}c^\dagger_{\r i\s}d^\dagger_{\r j\s'}d_{\r k\s'}c_{\r \l\s}.
\label{2ndband}
\eeq
The hopping parameters for the $c-$electrons are $\zeta_{\r\r'}$ and the chemical potential is denoted $\mu_c$. The couplings $V_{ijk\l}$ represent inter-species interaction matrix-elements. We will always be interested in the regime where the bandwidth of $c-$electrons is much bigger than the corresponding bandwidth of $d-$electrons, $\ie$ $W_c\gg W_d$---a situation that is often encountered in mixed-valence compounds \cite{MVrev}. 

At weak-coupling, $\ie$ when $U,~V\ll W_d,~W_c$, a simple perturbative approach is sufficient to conclude that the system is described by a Fermi liquid (for a generic density of $c,~d-$electrons) over a broad range of temperatures. On the other hand, the fate of the metallic state at strong-coupling, where a perturbative treatment is no longer justified, is unclear. Controlled theoretical progress, be it analytical or numerical, is impossible without imposing additional structure on the interaction matrix elements. In order to draw concrete conclusions for the above model in the regime of strong-coupling, we introduce a specific structure that makes the model `solvable' in the limit of large$-N$.

\subsection{A solvable limit}
\label{syksol}
Let us return to Eqs.~\ref{oneband},~\ref{2ndband} and treat the interaction matrix elements as independent Gaussian random variables. We further make the following assumptions:
\beq
&&\overline{U_{ijk\l}} = 0,~~\overline{U^2_{ijk\l}} = U^2,\label{choice}\\
&&U_{ijk\l} = -U_{jik\l} = -U_{ij\l k},~~U_{ijk\l}=U_{k\l ij},\\
&&\overline{V_{ijk\l}} = 0,~~\overline{V^2_{ijk\l}} = V^2,\\
&&V_{ijk\l} = -V_{jik\l} = -V_{ij\l k},~~V_{ijk\l}=V_{k\l ij},
\eeq
where `$\overline{^{~~~}}$' denotes `disorder' averaging. Notice that while the above choice implicitly assumes that we are averaging over different disorder realizations, each realization as defined by Eq.~\ref{oneband} and \ref{2ndband} has exact translational symmetry, $\ie$  we choose $U_{ijk\l},~V_{ijk\l}$ to be identical at every lattice site \cite{DCsyk}. At large$-N$, the model is self-averaging, so there is no distinction between properties of a single realization and disorder-averaged quantities.

What is the benefit of choosing such an artificial looking form for the interaction matrix elements? 
The astute reader will notice that by itself (and without the extra `spin-label'), the above form of the interaction Hamiltonian describes the well-known complex `Sachdev-Ye-Kitaev' (SYK) model \cite{SY,kitaev_talk}, which is a purely $(0+1)-$dimensional model for a strongly disordered interacting `quantum-dot'. At low-energies, the complex SYK model realizes a compressible, gapless phase \cite{SY,Parcollet2} without any long-lived quasiparticles. A number of recent works have studied the transport properties associated with higher dimensional lattice generalizations of SYK islands with strong disorder \cite{Gu17,SS17,Balents,Zhang17,SSmagneto,DVK17,hongyao,VS18a,VS18b,Altland2019}. The model defined in Eq.~\ref{oneband} and \ref{2ndband} is unique in that it has an exact translational symmetry for each realization of the interations~\cite{DCsyk} and an additional spin-label~\cite{DCEB}. These ingredients make it possible to address the fate of Fermi surfaces and their possible instabilities to superconductivity in the strong-coupling limit, as we discuss below.

\begin{figure}[t]
	\begin{center}
		\includegraphics[width=0.90\columnwidth]{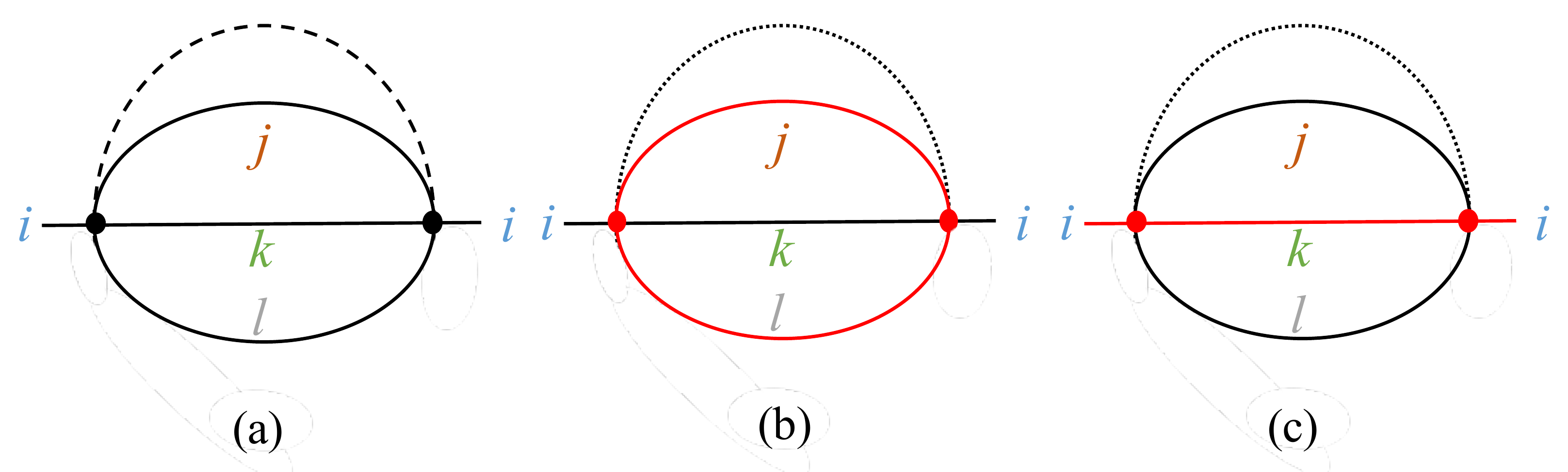}
	\end{center}
	\caption{The melonic diagrams with $O(U^2)$ contribution to (a) $\Sigma_d(\k,i\omega)$ and $O(V^2)$ contribution to (b) $\Sigma_d(\k,i\omega)$, (c) $\Sigma_c(\k,i\omega)$. Solid black (red) lines denote fully dressed $d-$ ($c-$) electron propagators. Dashed (dotted) lines denote $U^2$ ($V^2$) contractions.}
	\label{melon}
\end{figure}

Let us briefly review the structure of the large$-N$ saddle point equations, after averaging over the different realizations (each of which have perfect translational symmetry), which reduce to a set of self-consistent equations for the electron Green's function, $G_c(\k,i\omega),~G_d(\k,i\omega)$, and the interacting self-energies, $\Sigma_c(\k,i\omega),~\Sigma_d(\k,i\omega)$,
\beq
G_d(\k,i\omega) &=& \frac{1}{i\omega - \ve_\k - \Sigma_d(\k,i\omega)}, \label{saddle}\\
G_c(\k,i\omega) &=& \frac{1}{i\omega - \epsilon_\k - \Sigma_c(\k,i\omega)},\\
\Sigma_d(\k,i\omega) &=& -U^2\int_{\k_1}\int_{\omega_1} G_d(\k_1,i\omega_1)~\Pi_d(\k+\k_1,i\omega+i\omega_1)\nonumber\\
&& -V^2\int_{\k_1}\int_{\omega_1} G_d(\k_1,i\omega_1)~\Pi_c(\k+\k_1,i\omega+i\omega_1),\\
\Sigma_c(\k,i\omega) &=& -V^2\int_{\k_1}\int_{\omega_1} G_c(\k_1,i\omega_1)~\Pi_d(\k+\k_1,i\omega+i\omega_1),\\
\Pi_{c(d)}(\q,i\Omega) &=& \int_{\k}\int_{\omega} G_{c(d)}(\k,i\omega)~G_{c(d)}(\k+\q,i\omega+i\Omega).\label{eq:Pi_d}
\eeq
The dispersions for the $d$ and $c$ electrons are denoted $\ve_\k$ and $\epsilon_\k$, respectively. The equations above can be represented in a compact fashion diagramatically in terms of the `melonic' series in Fig.~\ref{melon}. These equations are  highly non-linear and seemingly complicated, and yet the beauty of these equations is that a simple description of NFL behavior emerges at the level of a single site at low energies. Such effectively `local' criticality arises when the temporal correlation functions have a power-law decay (up to correlation times $\xi_\tau\sim 1/T$), while the spatial correlations are exponentially decaying over a few lattice constants. We describe these solutions in the next section. 

\subsection{Marginal Fermi liquid from local criticality}
\label{mfl}

From a purely scaling point of view, the hopping terms are relevant compared to the SYK interaction terms. It is thus not surprising that the system at asymptotically low temperatures will be described by a Fermi liquid; remarkably the scale at which the crossover to the FL regime sets in for the $d-$electron system is $T_{\tn{coh}}\sim W_d^2/U$ \cite{Parcollet1,Balents,DCsyk}, which can be made parametrically small compared to the bare scales of $W_d$ and $U$. The primary issue we address in this and the next section is the nature of the metallic state above this crossover scale and its tendency towards various forms of electronic ordering (specifically towards pairing). 

Focusing on the $d-$electron system for now ($\ie$ setting $V=0$ and ignoring the decoupled free $c-$electrons), we now define the strong-coupling limit as: $W_d\ll U (\rightarrow\infty)$, with $T_{\tn{coh}}$ finite. In this limit, the solution for the Green's function upon solving Eq.~\ref{saddle}  has the form~\cite{DCsyk},
\begin{equation}
G_d(\k,i\omega) \sim 
\begin{cases}
\frac{Z}{i\omega - Z \overline\varepsilon_\k + i \gamma \nu_0^2 U|\omega|^2\ln\left(\frac{W^*}{|\omega|}\right)\mathrm{sgn}(\omega)}, ~\omega \ll W^*,\\
\frac{i\mathrm{sgn}(\omega)}{\sqrt{U |\omega|}} - B(\omega)\frac{\varepsilon_{\k}}{U |\omega|}, ~~~~~W^* \ll \omega \ll U,
\end{cases}
\label{limits}
\end{equation}
where $W^*\sim T_{\tn{coh}}\sim W_d^2/U$ is also the renormalized bandwidth for the $d-$electrons, $Z$ is the quasiparticle residue, $\overline\varepsilon_\k$ is the renormalized dispersion 
($\overline\varepsilon_\k / \varepsilon_\k$ is of order unity in the strong coupling limit), 
and $\gamma$ is a number of order unity (the $\log$ appears only in two-dimensions). The factor of $B(\omega)$, that descends from the ``spectral asymmetry" \cite{SS15}, is a constant independent of frequency but whose value depends only on the sign of $\omega$. 

There are a number of remarkable features associated with the above solution, which can be obtained analytically in two asymptotic regimes. As we already discussed, the appearance of a FL at low-energies, while interesting, is not entirely surprising. However, the FL is not a plain-vanilla metal; the quasiparticle weight is strongly renormalized as  $Z\sim 1/(\nu_0 U)$, where $\nu_0\sim1/W$ is the single-particle density of states (we use units where the lattice spacing $a=1$) and is accompanied by a strong mass-renormalization, $m^*/m\sim U/W (\sim Z^{-1}$). Within the above model, the FL has a dynamical mean-field theory (DMFT) like character \cite{DMFT}, where the frequency dependent renormalization of the self-energy is much larger than the associated momentum dependence. At high-energies, or, temperatures above $T_{\tn{coh}}$ (while still being small compared to $\ve_F$), the FL-like description is lost entirely. Approaching this scale from below, it is interesting to note that the single-particle scattering rate, $\Sigma_d''(\omega\rightarrow W^*)\sim W^*$, such that the notion of long-lived quasiparticles near the Fermi surface can no longer make sense. On the other hand, at weak-coupling ($U\ll W_d$), the system remains a FL at all temperatures with $Z\sim 1 - (\nu_0 U)^2$; all physical quantities can be obtained by carrying out perturbation theory in $U/W$ and the infinite resummation of the melonic diagrams becomes redundant.

The metallic regime at scales above $W^*$ is completely incoherent---there is no sharply defined surface in momentum space (resembling a Fermi surface) and there are no long-lived quasiparticles{\footnote{These notions can be made precise by taking the limit of $W^*\rightarrow0$ followed by $T\rightarrow0$.}}. This regime realizes a compressible NFL (where the compressibility scales as $1/U$) and ultimately controlled by the properties of the single SYK site, where the inter-site hoppings only enter perturbatively. The NFL obtained above is `infra-red incomplete' \cite{DCsyk} and is accompanied by a finite residual entropy, which is relieved below $W^*$ upon the crossover into the FL regime. 

At this point, we may ask if both of these metallic regimes are stable against pairing (and possibly other) instabilities, or if all of the interesting NFL properties are masked by such ordering tendencies? This will be the primary subject of discussion in the following section. But before we do so, let us investigate the possible role played by the $c-$electrons once they are coupled to the strongly interacting $d-$electrons by turning on a finite $V$. 

It is clear that when the strongly renormalized heavy FL is coupled to the non-interacting $c-$electrons at scales below $T_{\tn{coh}}$, the resulting state remains a FL with two independently conserved densities. On the other hand, when the $c-$electrons scatter off the fluctuations associated with the SYK-like $d-$electrons, the self-energy becomes,
\beq
\Sigma_c(i\omega) = -\frac{\nu_0 V^2}{2\pi^2U} i\omega\log\bigg(\frac{U}{|\omega|}\bigg).
\label{eq:sigma_mfl}
\eeq
This is the celebrated marginal Fermi liquid (MFL) form of the self-energy \cite{Varma}, which arises purely as a result of scattering off an effective `bath' {\footnote{Though it is important to note that there is no real `bath' as the number of degrees of freedom in the two systems is comparable.}} formed by the incoherent SYK-like $d-$electrons . The $c-$electrons have a sharply defined Fermi surface{\footnote{Formally defined as the solution to $G_c^{-1}(\k,\omega=0)=0$ as $W^*\rightarrow0$ followed by $T\rightarrow0$.}} that also satisfies Luttinger's theorem \cite{DCsyk}. However, the nature of the critical Fermi surface that arises in the above model is significantly different from the one that arises from coupling to $\eg$ the gapless nematic order parameter (as discussed in Sec.~\ref{double}). The key difference is in the structure of the momentum dependence of the correlation functions. While the singular frequency dependence of the self-energy is restricted to the near vicinity of the critical Fermi surface for the problem with a gapless boson, this is not the case for the SYK model. In the latter, the frequency dependence has the singular structure everywhere in momentum space, even away from the Fermi surface---a feature that arises from the local SYK island. Let us now address the central question of this review: Is the critical Fermi surface obtained within the above solvable model unstable to pairing or other competing instabilities?

\subsection{Strong-coupling superconductivity and intertwined orders}
\label{SCsyk}

A number of recent works \cite{CX17,Patel,YW19,esterlis,DCEB} have studied the nature of pairing instabilities in variants of SYK-type models. In particular, most of these studies have focused on the single-site ($\ie$ 0-dimensional) SYK model (and extensions thereof)---either by explicitly including attractive interactions \cite{Patel} or by coupling to a gapless boson \cite{YW19,esterlis}---and studied the resulting pairing instabilities to have a non-BCS form. The present authors studied the pairing instabilities of the single-band model ($\ie$ with only $d-$electrons) with spatial structure \cite{DCEB}, which we review below and also discuss some new results on the two-band generalization. Quite remarkably, for all of the above studies, the pairing instabilities in the regime where the analysis is controlled can be described within Eliashberg theory. 

\begin{figure}[t]
	\begin{center}
		\includegraphics[width=0.90\columnwidth]{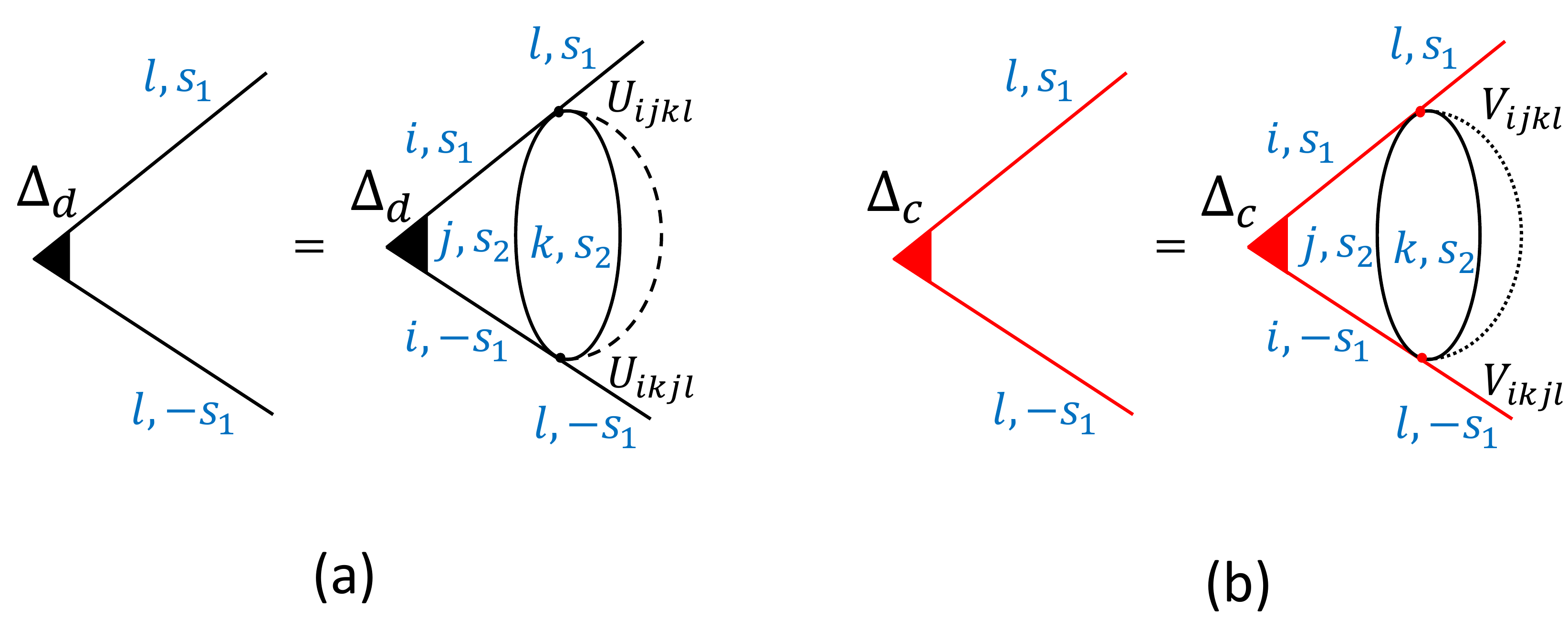}
	\end{center}
	\caption{The linearized gap equations for the pairing vertices for (a) $\Delta_d(\k,i\omega)$ and (b) $\Delta_c(\k,i\omega)$. Different lines are identical to those introduced in Fig.~\ref{melon}.}
	\label{SYKeliash}
\end{figure}

Let us first address the instabilities for the model with just $d-$electrons ($\ie$ setting $V=0$). Then, in addition to the structure imposed on the $U$'s in Eq.~\ref{choice}, we now impose the following constraints{\footnote{Note that it is also possible to consider these to be statistically independent, in which case there are no solutions to the linearized gap equation to leading order in $1/N$.}}:
\beq
U_{ijk\l} = \pm U_{ikj\l}.
\label{Ucons}
\eeq
Imposing this constraint does not modify the solution for the single-particle Green's function (Eq.~\ref{limits}). The linearized gap equations in the spin-singlet orbital diagonal pairing channel, that originate from the full Eliashberg equations (valid to leading order in $1/N$), take the simple form \cite{DCEB},
\beq
\Delta_d(\k,i\omega) = \pm U^2T\sum_{i\Omega}\int_\p \Delta_d(\p,i\Omega) G_d(\p,i\Omega)G_d(-\p,-i\Omega)\Pi_d(\k-\p,i\omega-i\Omega),
\eeq
which can also be expressed diagramatically as in Fig.~\ref{SYKeliash}a. Then, one finds two families of solutions, depending on the sign in Eq.~\ref{Ucons} above. 

When $U_{ijk\l}=+U_{ikj\l}$, the NFL regime at $T>T_{\tn{coh}}$ is unstable to on-site $s-$wave pairing with a finite $T_c$. The momentum dependence in the above equation can then be ignored as the physics is completely local and emerges at the level of a single site. Moreover, since there is only one scale in the problem, $T_c\sim U$ (with an $O(1)$ prefactor). The interesting sign structure of the matrix elements essentially generates attraction in the pairing channel, thereby masking the onset of any NFL regime. The spin-dependent structure of the model here is crucial --- a spinless version of the same problem would not show the same phenomenology, since (i) there is no on-site orbital diagonal `triplet' order that can be similarly defined, and, (ii) a true transition at $T_c$ (with a diverging susceptibility) in the spatially extended spin-triplet channel can not emerge due to a finite hopping ($t_{\r\r'}$), as long as the latter can be treated perturbatively.    

On the other hand, when $U_{ijk\l}=-U_{ikj\l}$, the NFL regime is stable against pairing all the way down to $T_{\tn{coh}}$, even though the pairing susceptibility grows in strength (without diverging) 
as a function of decreasing temperature. However, once there is a crossover into the incipient heavy FL regime, a generalized Kohn-Luttinger type mechanism \cite{KL,SR11} arising purely from the momentum dependence of the density-density correlation function guarantees a finite $T_c$. Interestingly, since $W^*$ (the renormalized bandwidth) is the only scale left over in the problem, $T_c\sim W^*$ (with an $O(1)$ prefactor). Moreover, this is a regime where the Fermi surface associated with the incipient FL hasn't fully formed, so the instability is not controlled by the sharpness of the Fermi surface. 

The linearized self-consistent equations for the instabilities in the particle-hole channel, describing density-wave instabilities, also have a solution in this regime \cite{DCEB}. Since the instabilities are not tied to any underlying `fermiology'---one is equally likely to find a tendency towards such forms of broken symmetries due to the same interactions, with comparable transition temperatures to that of the superconducting $T_c${\footnote{The precise ordering temperatures are determined by the numerical prefactors, which are controlled by the microscopic details of the specific model.}}.  The models introduced here are thus ideal playgrounds to study the phenomenon of {\it intertwined} orders, $\ie$, the occurence of different types of distinct, symmetry-unrelated forms of electronic order in the phase diagram, all arising from  the same microscopic interactions and with comparable onset temperatures. This phenomenon is of great interest for a description of the experimental phenomenology across numerous families of high-temperature superconductors \cite{RMPintertwined}.  

Finally, consider the pairing instabilities for the two-band model upon turning on a finite $V$. Assume the following additional structure on the interaction matrix elements,
\beq
V_{ijk\l} = \pm V_{ikj\l},
\label{vcons}
\eeq
and, unlike Eq.~\ref{Ucons}, treat $U_{ijk\l}$  and $U_{ikj\l}$ as statistically independent (we do this in order to avoid any intrinsic SC instability associated with the $d-$electrons). The linearized gap equations in the spin-singlet orbital diagonal pairing channel for the $c-$electrons takes the form,
\beq
\Delta_c(\k,i\omega) = \pm V^2 T\sum_{i\Omega} \int_\p \Delta_c(\p,i\Omega)G_c(\p,i\Omega)G_c(-\p,-i\Omega)\Pi_d(\k-\p,i\omega-i\Omega).
\label{eq:BS_marginal}
\eeq
Let us first focus on the case where there is a positive sign in Eq.~\ref{vcons}, in which we can get a non-trivial solution to Eq.~\ref{eq:BS_marginal} even neglecting the (weak) momentum dependence of $\Pi_d$. After integrating over the momentum perpendicular to the Fermi surface, we obtain the following equation for the pairing vertex:
\beq
\Delta_c(i\omega)=\pi \nu_0 V^2T  \sum_{i\Omega} \Delta_c(i\Omega) \frac{\Pi_d(i\omega - i\Omega)}{|i\Omega -\Sigma_c(i\Omega)|}.
\label{eq:mfl1}
\eeq 
Here, $\Sigma_c(i\omega)$ is given by Eq.~\ref{eq:sigma_mfl}, and $\Pi_d$ has the SYK form: $\Pi_d(i\omega) \sim \frac{1}{U}\log\left(\frac{U}{|\omega|}\right)$.  

It is instructive to examine the limit $V\ll W_c\sim U$. In this limit, we can show that the superconducting $T_c$, where a solution to Eq.~\ref{eq:mfl1} first appears, is parametrically larger than the marginal Fermi liquid scale $\Omega_{\mathrm{MFL}}$, defined as the energy scale at which $|\Sigma_c(i\Omega_{\mathrm{MFL}})| = \Omega_{\mathrm{MFL}}$. This follows from an analysis along the lines of Ref.~\cite{DCEB}, the details\footnote{In this case, Eq.~\ref{eq:mfl1} is essentially identical to Eq.~9 of Ref.~\cite{DCEB}.} of which will be presented elsewhere \cite{unpub}; the result is $T_c\sim U e^{-\sqrt{\frac{U}{\nu_0 V^2}}}$, compared to $\Omega_{\mathrm{MFL}}\sim U e^{-\frac{U}{\nu_0 V^2}}$. In this limit, therefore, the marginal Fermi liquid regime is preempted by superconductivity. In the intermediate coupling case, $V\sim U\sim W_c$, both $\Omega_{\mathrm{MFL}}$ and $T_c$ are of the order of $W_c$, and there is no parametrically broad MFL regime. 

In contrast, for the case of a negative sign in Eq.~\ref{vcons}, there is no solution to Eq.~\ref{eq:BS_marginal} if the polarizability $\Pi_d$ is momentum independent. In the limit $W_d\rightarrow 0$, $\Pi_d$ is completely momentum-independent to leading order in $1/N$ ($\Pi_d$ is a convolution of the two $d$ electron Green's functions, Eq.~\ref{eq:Pi_d}), and for $W_d=0$, $G_d$ is completely momentum independent\footnote{This follows from the fact that when $W_d=0$, the  number of $d$ electrons is conserved on every site seperately.}. Hence there is no pairing instability in this limit, and the marginal Fermi liquid regime is stable at order $1/N$. If $W_d\ne 0$ then $\Pi_d$ has momentum dependence derived from that of $G_d$ in Eq.~\ref{limits}. Then, it is possible to get non-trivial momentum dependent solution to Eq.~\ref{eq:BS_marginal}. Nonetheless, in the limit where $W_d\ll  U$, $T_c$ will be parametrically smaller than $T_{\mathrm{MFL}}$, and there is a parametrically  broad marginal Fermi liquid regime.

\section{Summary and Outlook}
\label{outlook}
Migdal-Eliashberg theory was designed to treat the electron-phonon problem, where the separation of scales between the Fermi energy and the Debye frequency provides a small parameter. Unfortunately, in strongly correlated electronic systems, there is no such natural small parameter that allows theoretical control. Finding special controlled limits --- even artificial ones --- could provide a useful handle on the rich non-perturbative physics that may emerge in these systems. In this article, we have reviewed two such controlled limits: the double expansion in small $\epsilon$ and $1/N$ for the problem of a Fermi surface coupled to a gapless boson, and the large $N$ limit for the problem of lattice fermions with $N-$orbitals and frustrated interactions. Interestingly, in both limits, Eliashberg theory becomes exact, and describes the competition between the destruction of Landau quasiparticles and the pairing tendency, both effects arising from the same underlying mechanism. 

It may well be that, in many situations, Eliashberg theory is valid far beyond its formal range of validity\footnote{As the physicists' saying goes, a good theory is one that works at least an order of magnitude beyond its range of validity.}. Quite interestingly, in some of the examples where Eliashberg theory appears to work well without any formal justification \cite{EB17}, it is unclear what is responsible for this agreement. This is not to say, however, that all non-Fermi liquids are well-described by Eliashberg theory; for instance, a recent work revealed a strongly coupled fixed point of a nearly antiferromagnetic metal, with a very different structure~\cite{Schlief2017}, which is controlled by an {\it emergent} small parameter. The search for such new non-perturbative fixed points, as well as other controlled limits of correlated electron systems, is bound to provide new insights into the complexity of the growing class of correlated quantum materials.   

In all of the examples discussed in this review where Eliashberg theory is applicable, the superconducting transition is fundamentally described within a generalized `mean-field' type approach. In settings where the interactions are much larger than the bandwidth and the superfluid density is small, fluctuation effects near the superconducting transition are bound to be significant and a mean-field description is likely to fail. The recent discovery of superconductivity in a number of two-dimensional graphene based Moir\'{e} superlattices ~\cite{Cao2018,AY19,Efetov,FW19,PK19} provide an ideal platform to investigate such effects. All of these materials are believed to host isolated, nearly flat topological bands where interactions are comparable (or possibly, even larger) than the free electron bandwidth. Inspired by these rapid developments, one is tempted to ask the following question: What is the highest possible $T_c$ for an isolated flat band in the limit of its bandwidth going to zero when the interaction strength is finite? In the absence of any small parameter, when the interaction strength sets the only scale in the problem and the carrier density is low, it is unclear if an Eliashberg type approach can capture the superconducting instabilities, if any. Developing a controlled analytical framework, that goes beyond the Migdal-Eliashberg formalism, to investigate the onset of pairing instabilities in such flat-band systems remains an interesting challenge for the future.
\\

{\bf Acknowledgements}
We thank M. Metlitski and T. Senthil for helpful discussions. DC is supported by startup funds at Cornell University. EB is supported by the European Research Council (ERC) under grant HQMAT (grant no. 817799), by the Israel-USA Binational Science Foundation (BSF), and by the Minerva foundation.

\bibliography{mybibfile}

\end{document}